\documentclass[aps,pre,superscriptaddress,noshowpacs]{revtex4}
\usepackage[intlimits]{amsmath}
\pdfoutput=1
\usepackage{amsfonts}
\usepackage{psfrag}
\usepackage{graphicx}
\usepackage{marvosym}
\usepackage{subfigure}
\usepackage[usenames]{color}
\usepackage{indentfirst}
\usepackage[utf8]{inputenc}
\usepackage[T1]{fontenc}
\usepackage{ifpdf}

\ifpdf
\usepackage{epstopdf}
\usepackage[pdftex]{hyperref}
\else
\usepackage[hypertex,ps2pdf,colorlinks,urlcolor=blue,citecolor=blue]{hyperref}
\fi

\advance\voffset by -1.4cm
\advance\hoffset by -0.3cm
\begin{document}

\title{Conformal invariance in three dimensional percolation}

\author{Giacomo Gori}
\affiliation{CNR-IOM DEMOCRITOS Simulation Center, 
Via Bonomea 265 I-34136 Trieste, Italy}

\author{Andrea Trombettoni}
\affiliation{CNR-IOM DEMOCRITOS Simulation Center, 
Via Bonomea 265 I-34136 Trieste, Italy}
\affiliation{SISSA, Via Bonomea 265 I-34136 Trieste, Italy}
\affiliation{INFN, Sezione di Trieste, I-34127 Trieste, Italy}

\begin{abstract}
The aim of the paper is to present numerical
results supporting the presence of conformal
invariance in three dimensional statistical
mechanics models at criticality and to
elucidate the geometric aspects of universality.
As a case study we study three dimensional percolation
at criticality in bounded domains.
Both on discrete and continuous models
of critical percolation, we test by numerical experiments the invariance
of quantities in finite domains under conformal
transformations focusing on crossing
probabilities. Our results show clear
evidence of the onset of conformal invariance
in finite realizations especially for
the continuum percolation models.
Finally we propose a simple analytical
function approximating the crossing probability
among two spherical caps on the
surface of a sphere and confront it
with the numerical results.

\end{abstract}
\maketitle

\section{Introduction}
\label{introduction}
A major fact in the theory of critical 
phenomena is that the scale invariance exhibited by systems at criticality \cite{mussardo_2010}
when combined with a local action may give rise 
to invariance under the larger group
of conformal transformation which locally acts 
as scale transformation \cite{di_francesco_1997}. Although being scale invariant and
having a local action is not in general equivalent 
to being conformal invariant, since there are scale invariant
local models which are not conformal invariant \cite{riva_cardy_2005}
and non-local models possessing conformal invariance 
\cite{witten_1998,heemskerk_2009,dorigoni_2009,rajabpour_2011}, 
it is widely believed to hold true for many models
of physical interest.

The conformal group in $d$ spatial dimensions (for $d\neq2$) 
has a number of independent generators
equal to $\frac{1}{2}(d+1)(d+2)$.
For the special case $d=2$ the conformal
group turns out to be infinitely dimensional \cite{di_francesco_1997}.
This infinite dimensionality is indeed at the
root of the success of conformal field
theories in the study of two dimensional critical models.
For higher dimensions the exploitation
of conformal symmetry (if present) leads nonetheless
to important simplifications in the study of critical
models. Even though for a long lapse of time traditional 
techniques not explicitly incorporating conformal 
invariance were used for the study of critical
phenomena (such as the $\epsilon$-expansion 
\cite{guida_1998}), 
the last years witnessed the development in three dimensions 
of the so called conformal bootstrap program which, fully
exploiting conformal invariance, has lead to high-precision nonperturbative predictions 
for anomalous scaling dimensions of the Ising model in three
dimensions \cite{rychkov_2012} and in fractional dimensions \cite{rychkov_2014}. The role of conformal symmetry
in the theory of critical phenomena
can thus be hardly underestimated. 

Dating back to the Polyakov's hypothesis of the conformal invariance of critical fluctuations \cite{polyakov_1970} 
and to the derivation under broad conditions in $d=2$ of (conformal) invariance 
under local changes of length scale from the (scale) invariance under rigid length scale changes
\cite{polchinski_88}, a significant amount of work was by then devoted to show with mostly analytical  
different techniques the possible existence of conformal symmetry at criticality in higher dimensions and its relation 
with the scale invariance. It was soon pointed out that unitarity is a key ingredient, with supersymmetric 
models in $d=4$ being an important example in which the conformal invariance was explicitly shown \cite{seiberg_1994}. 
Always in $d=4$ a proof of conformal invariance within perturbation theory has been recently discussed 
\cite{luty_2013,fortin_2013}. In $d=3$ we mention the recent discussion 
of the occurrence of conformal
invariance for the Ising model at criticality using functional renormalization 
group techniques \cite{delamotte_2015}.

For two dimensional systems, complementary to analytical tools and techniques, 
an important role in elucidating the geometric aspects 
of universality and in testing the emergence and consequences of conformal invariance 
has been played by numerical simulations:  
in $d=2$ conformal symmetry has been extensively
tested numerically (see e.g. the early numerical
experiments in \cite{cardy_1994} and \cite{langlands_1994}) by confronting numerics with theoretical predictions
arising from conformal field theory. 

However, in other space dimensionlities these 
tests and numerical experiments are still partial. 
An active research line dates back to the
paper \cite{cardy_1985}, in which
a relation among correlation
functions defined on a flat $d$-dimensional
geometry and a suitably chosen
curved geometry (the hypercylinder
$S^{d-1}\times \mathbb{R}$) has been put forward, which 
in $d=2$ amounts to
the conformal mapping between the (both
flat) plane and cylinder geometries. This relation was originally verified for the spherical model for
dimension $2<d<4$ \cite{cardy_1985}.
Recent Monte Carlo numerical studies
established the validity of this prediction
for the three dimensional Ising
model \cite{deng_2002, deng_2003}: in these works  
a continuous limit of the lattice model was used to 
perform the simulations of the curved space and the assumption of conformal invariance, 
together with the determination of the magnetic and energy correlation lengths, 
was shown to be enough to correctly reproduce the known critical exponents. Similar numerical 
computations were carried out for the three dimensional bond 
percolation \cite{deng_2004} and $O(N)$ models \cite{janke_2002}.

The aim of this paper is to provide numerical evidence
of the invariance of three dimensional critical systems under conformal transformations of the euclidean flat space (more precisely, of the flat space into itself). 
For this reason we studied observables of three dimensional percolation at criticality, studying models both on lattice (site percolation) and on the 
continuum (overlapping spheres). 
We decided to choose percolation for several reasons: 
percolation is possibly the simplest
model exhibiting a phase transition,
see e.g. \cite{stauffer_1992} and the recent review \cite{saberi_2015}. This simplicity has on one side
allowed to mathematically prove
many conjectured properties of percolation
such as the existence of a transition
in the thermodynamic limit \cite{aizenman_1987} or, more
recently, to prove rigorously \cite{smirnov_2006} many
results from CFT with help of  
of Stochastic Loewner Evolution (SLE) 
techniques \cite{schramm_2001}. On the other hand, with the aid of 
powerful algorithms \cite{newman_2000, newman_2001}, the numerical
exploration of percolation problems is
especially convenient. Finally, it is clear that percolation 
is an ideal system to test conformal invariance in bounded 
domains, as convincingly showed for $d=2$ \cite{langlands_1994}, 
since in a natural way one computes quantities defined and crucially dependent 
on the form of the boundaries, as the probability of connecting two disconnected 
regions of the boundary of the system. 

With these motivations, we numerically study the presence of conformal
invariance for three dimensional percolation
at criticality in bounded domains.
We introduce and study both discrete and continuous models
of critical percolation, we test by numerical experiments the invariance
of quantities in finite domains under conformal
transformations by computing crossing
probabilities. We also put forward
an approximate expression
for the crossing among spherical caps
on a sphere and check it against numerical
experiments finding very good agreement.
Our results show clear
evidence of the onset of conformal invariance
in finite realizations especially for
the continuum percolation models.

\section{Percolation models and conformal invariance}\label{percolation}

In its simplest incarnation percolation consists
of filling a randomly and independently
with probability $p$ the sites of a graph 
and checking for the existence of large clusters.
As a testbed for the verification of
conformal invariance we will consider
a classical problem in percolation
theory: the existence of a 
cluster connecting two disjoint
regions $\omega_1$ and $\omega_2$ of 
the boundary of a
bounded domain $\Omega$. The probability
of the occurrence of this event
is known (and under some assumptions 
mathematically proven \cite{aizenman_1987}) to tend,
in the thermodynamic limit, to
a nontrivial (i.e. different from $0$ or $1$) value $\pi_{\times}$  
if the occupation probability $p$ is set to the 
critical probability $p_c$.
Before introducing the finite realizations
considered in this work (this will be done in
Section \ref{percolation_models}) we 
explore in this Section the geometric consequences of 
conformal invariance by working in the 
thermodynamic limit.
Due to universality these considerations
do not depend on the specific model
of percolation under scrutiny.
In order to verify this
model independence we will
take two prototypical models
of percolation: a discrete and
a continuum one, i.e. site percolation
and continuum percolation of penetrable
spheres, with the notations introduced in this Section being applicable 
to both models.

The crossing probability has been the
object of classical foundational 
papers \cite{cardy_1994, langlands_1994} 
for the application of conformal field 
theory in two dimensional critical systems.
In the two dimensional case indeed full
exploitation of the (infinite dimensional)
conformal group has allowed to obtain
the exact crossing probability for simply
connected domains in the 
paper by Cardy \cite{cardy_1994}.
In that setting the presence of 
boundaries gets translated, in the
boundary CFT language, into the
insertion of boundary creating operators
located at the points at the extremes of the
one dimensional boundary.
Although many similar formulas
in different and more complicated settings
have been derived \cite{watts_1996}, some details
of the operator content
in percolation in two dimensions,
which turns out to be a logarithmic
CFT, are still matter of investigation
\cite{simmons_2013}.

In our three dimensional setting
the domain $\Omega$ will be chosen as a
simply connected closed domain 
and the two surface domains as
intersection of the boundary
$\partial \Omega$
and two closed domains $\Omega_1$ and $\Omega_2$.
Although the two domains $\Omega_1$ and $\Omega_2$ 
are not strictly needed to define  of the regions 
$\omega_1$ $\omega_2$
they will come in handy to define 
unambiguously the finite lattice realization
we will study numerically.

At criticality the crossing probability
should take a value $\pi_{\times}(\Omega, \Omega_1, \Omega_2)$
dependent on the three domains
$\Omega$, $\Omega_1$ and $\Omega_2$.
Please note that the actual value of
$\pi_{\times}(\Omega, \Omega_1, \Omega_2)$
\emph{does not} depend on the percolation
model we are studying.
This expectation has been numerically
validated in widely different 
models in the two dimensional models
and in some selected three dimensional
cases \cite{skvor_2007, lorenz_1998}.

If we perform a conformal
mapping $\mathfrak{C}$ the three
domains get mapped onto the three domains
$\mathfrak{C}(\Omega)$, $\mathfrak{C}(\Omega_1)$ and 
$\mathfrak{C}(\Omega_2)$.
The conformal invariance hypothesis
implies that:
\begin{equation}
 \pi_{\times}(\Omega,\Omega_1, \Omega_2) = \pi_{\times}(\mathfrak{C}(\Omega), \mathfrak{C}(\Omega_1), 
\mathfrak{C}(\Omega_2)).\label{CFT_inv}
\end{equation}
Note that in general conformal invariance would allow
for dimensionful prefactors in front of
\eqref{CFT_inv} depending
on the geometry and on the scaling
dimension of the observable under
consideration. The further requirement
on the crossing probability to be,
as expected, a finite number in the thermodynamic limit
forces us to set them to one.
In the two dimensional case, using 
the CFT language, the above statement
can be expressed by saying that the
boundary operators we insert to
obtain the crossing probability are expected to have zero
conformal weight.
In three dimensions of course the interpretation of the
boundary condition as insertion of boundary operators is lacking
but the expectation of the crossing probability to have 
vanishing scaling dimension still holds.

Another observation is in order
the above statements on implication 
of conformal invariance crucially
depend on the choice of 
the boundary conditions chosen.
Indeed 
different boundary conditions
can flow to \cite{diehl_1998, pleimling_2004}
renormalization group fixed points differing
from the one encountered in
the bulk systems. A study based on $\epsilon$-expansion of the effect of boundary conditions 
on the critical properties of percolation models in semi-infinite geometries has been 
carried out in \cite{diehl_1989}. The boundary conditions we have chosen
i.e. occupied for points in
$\omega_i$ and free for the remaining
part of the boundary are expected, in analogy
to what observed in two dimension, to
the so called ordinary fixed point.

We recall
that in three dimensional
space the conformal group is
10-dimensional \cite{di_francesco_1997} and is generated by
the following generators
(acting on a three dimensional
vector $\mathbf{x}$):

\begin{itemize}
 \item 3 translations: $\partial_\mu$
 \item 1 dilatation: $\sum_\mu x_\mu \partial_\mu$
 \item 3 rotations: $x_\mu \partial_\nu - x_\nu \partial_\mu$
 \item 3 special conformal transformations: $\sum_\nu (2 x_\mu x_\nu \partial_\nu 
- x_\nu x_\nu \partial_\mu)$
\end{itemize}
where $x_\mu$ denote
the components of $\mathbf{x}$. 
Indeed the special conformal
transformations are most interesting ones
and we will concentrate on them.

We will consider different starting geometries
and then examine the problem obtained by
applying compositions of the following
conformal transformations:

\begin{equation}
x'_\mu = \frac{(1-v^2)(x_\mu+ v_\mu)+v_\mu (\mathbf{x}+\mathbf{v})\cdot(\mathbf{x}+\mathbf{v})}{1+2\mathbf{x}\cdot\mathbf{v}+x^2 v^2}
\label{transformation}
\end{equation}
where the vector $\mathbf{v}$ rules the 
magnitude and direction of the transformation.
We will concentrate on the two transformations
$\mathfrak{C}_x$ and $\mathfrak{C}_z$ having
vectors $\mathbf{v}=\varepsilon\{1,0,0\}$ and 
$\mathbf{v}=\varepsilon\{0,0,1\}$ respectively
and compositions of them.  
The parameter $\varepsilon$ ruling the magnitude
of the transformation (when $\varepsilon=0$
transformation \eqref{transformation} reduces 
to the identity) 
will be set conventionally to $0.2$.
These transformations, which are 
finite exponentiations of the generators 
$- \partial_\mu + \sum_\nu (2 x_\mu x_\nu \partial_\nu 
- x_\nu x_\nu \partial_\mu)$, i.e. combination
of translations and special conformal transforms, 
have the property of mapping the unit sphere onto
itself.
In table \ref{transformation_table} we write
the transformation which will
be considered together with the roman
numeral used to denote it.

\begin{table}
\centering
\begin{tabular}{|l|l|}
  \hline
  numbering & transformation\\ \hline
  I & $\mathbf{1}$ (identity) \\
  II & $\mathfrak{C}_z$ \\
  III & $\mathfrak{C}_x$ \\
  IV & $\mathfrak{C}_z \mathfrak{C}_x$ \\
  V & $\mathfrak{C}_x \mathfrak{C}_z$ \\
  VI & $\mathfrak{C}_z \mathfrak{C}_z$ \\
  VII & $\mathfrak{C}_x \mathfrak{C}_x$ \\
  VIII & $\mathfrak{C}_z \mathfrak{C}_z \mathfrak{C}_z$ \\
  IX & $\mathfrak{C}_x \mathfrak{C}_x \mathfrak{C}_x$ \\
  \hline
\end{tabular}
\caption{Definition of the considered conformal transformations.}\label{transformation_table}
\end{table}

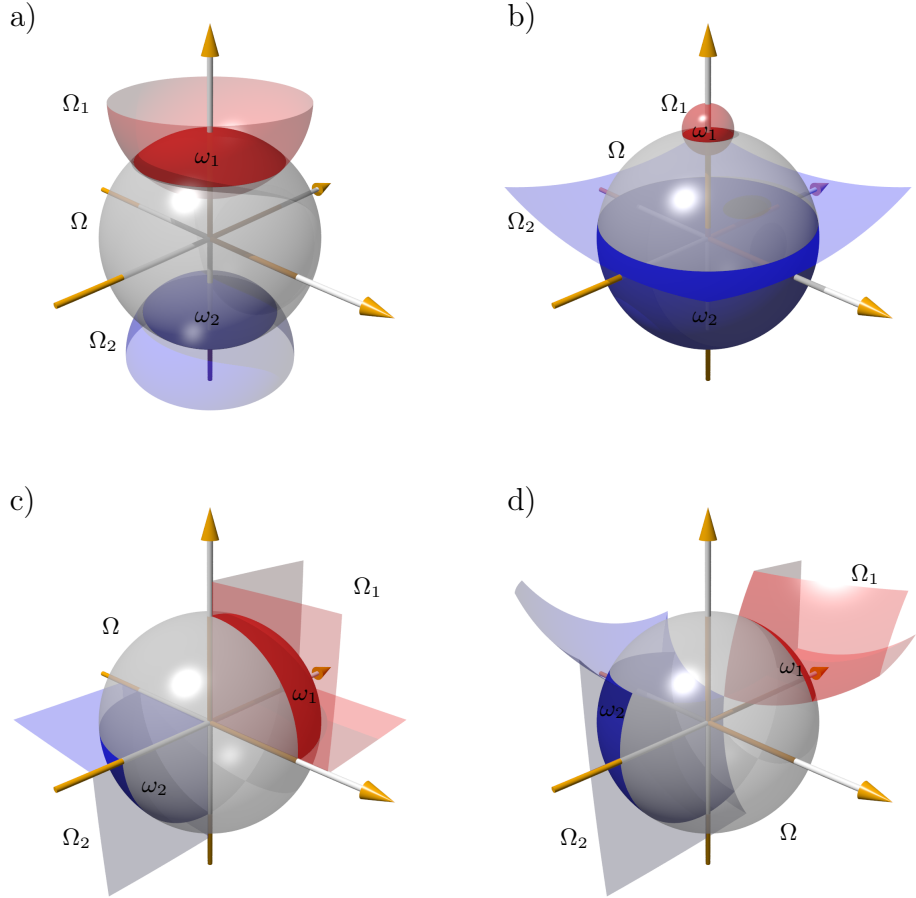
\begin{figure}
\centering
 \includegraphics[width=.4\textwidth]{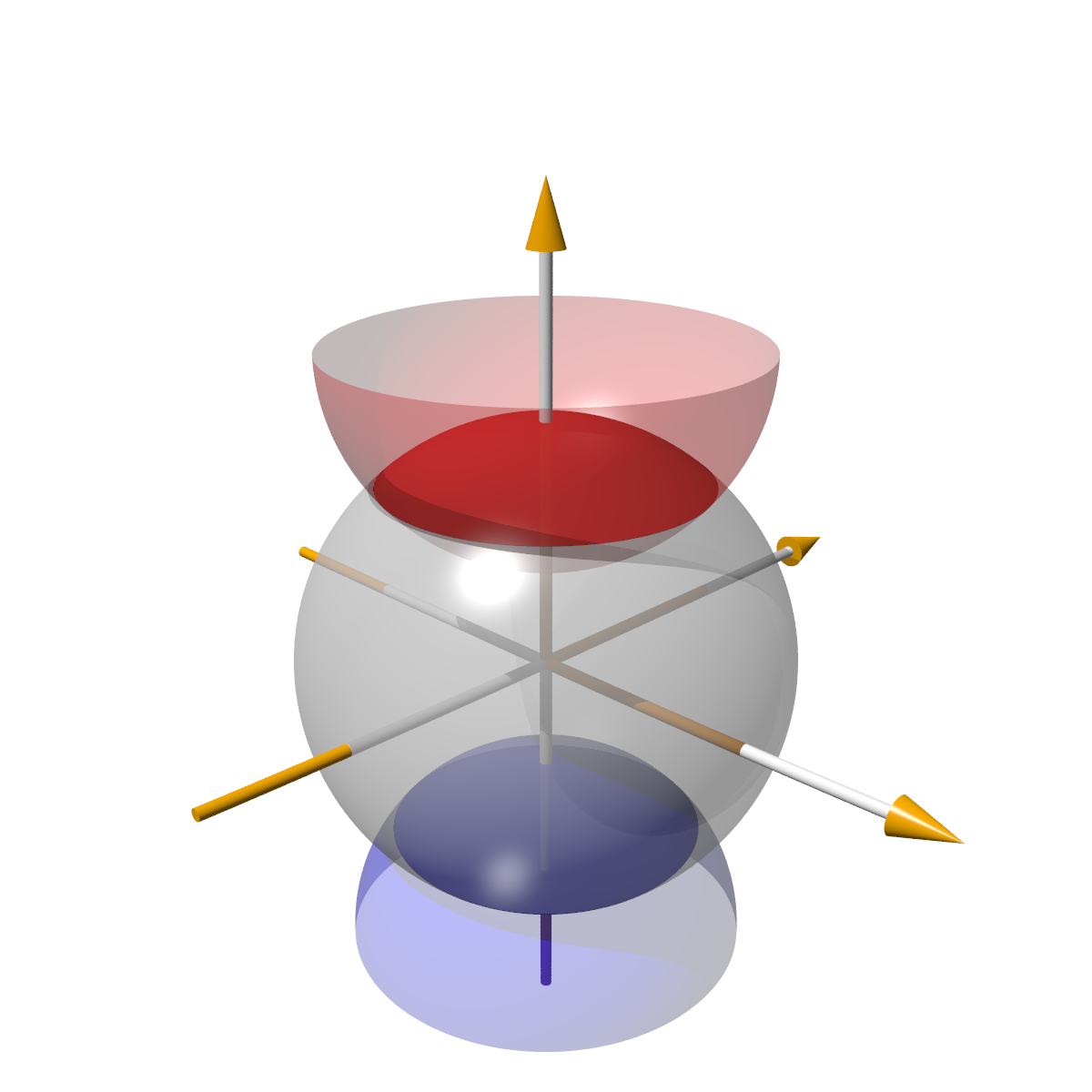}
\begin{picture}(0,0)
\put(-170,152){{\large a)}}
\put(-150,120){${\large \Omega_1}$}
\put(-147,75){${\large \Omega}$}
\put(-140,30){${\large \Omega_2}$}
\put(-100,100){${\large \omega_1}$}
\put(-100,40){${\large \omega_2}$}
\end{picture}
 \includegraphics[width=.4\textwidth]{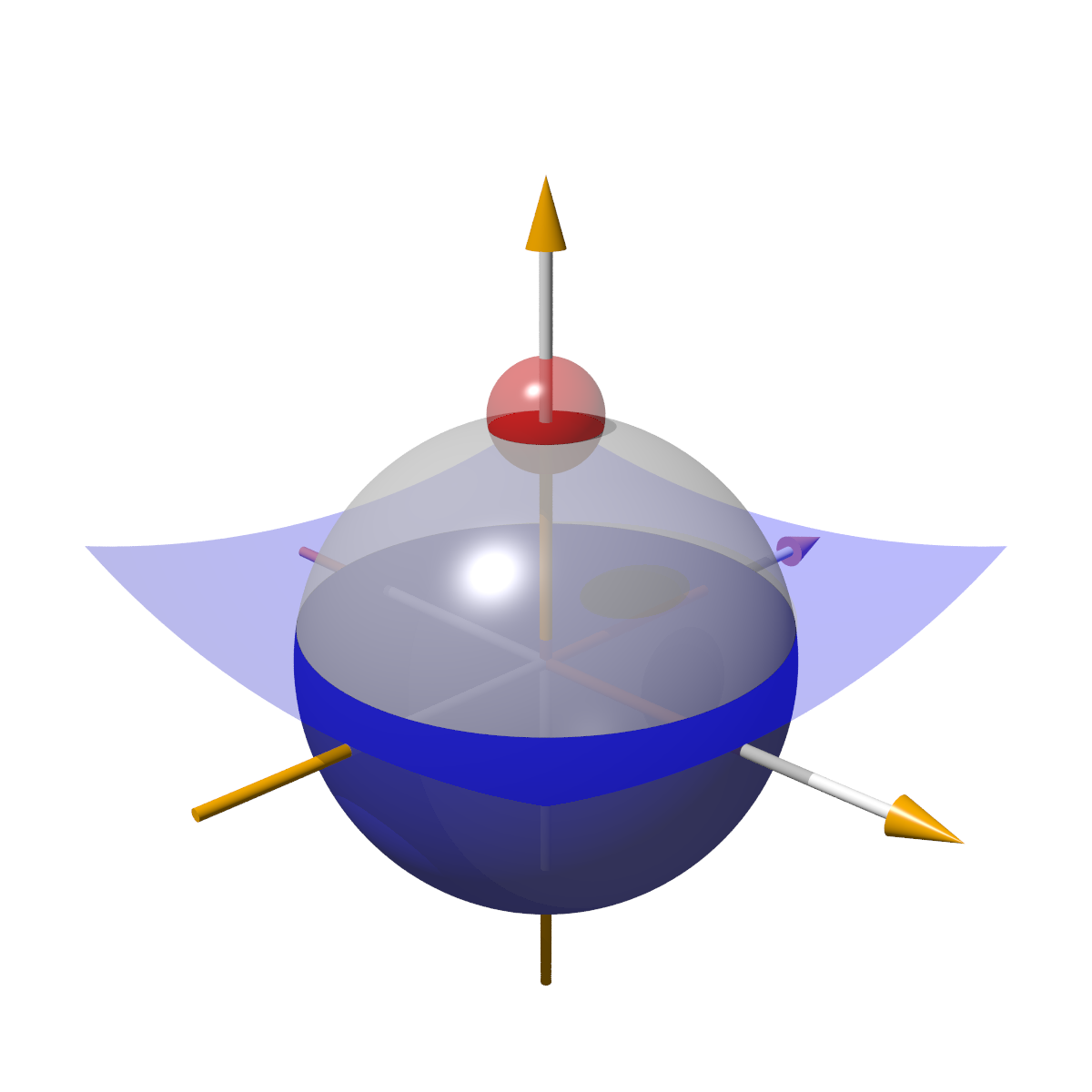}
\begin{picture}(0,0)
\put(-170,152){{\large b)}}
\put(-112,120){${\large \Omega_1}$}
\put(-132,102){${\large \Omega}$}
\put(-170,75){${\large \Omega_2}$}
\put(-100,110){${\large \omega_1}$}
\put(-100,40){${\large \omega_2}$}
\end{picture}
 \includegraphics[width=.4\textwidth]{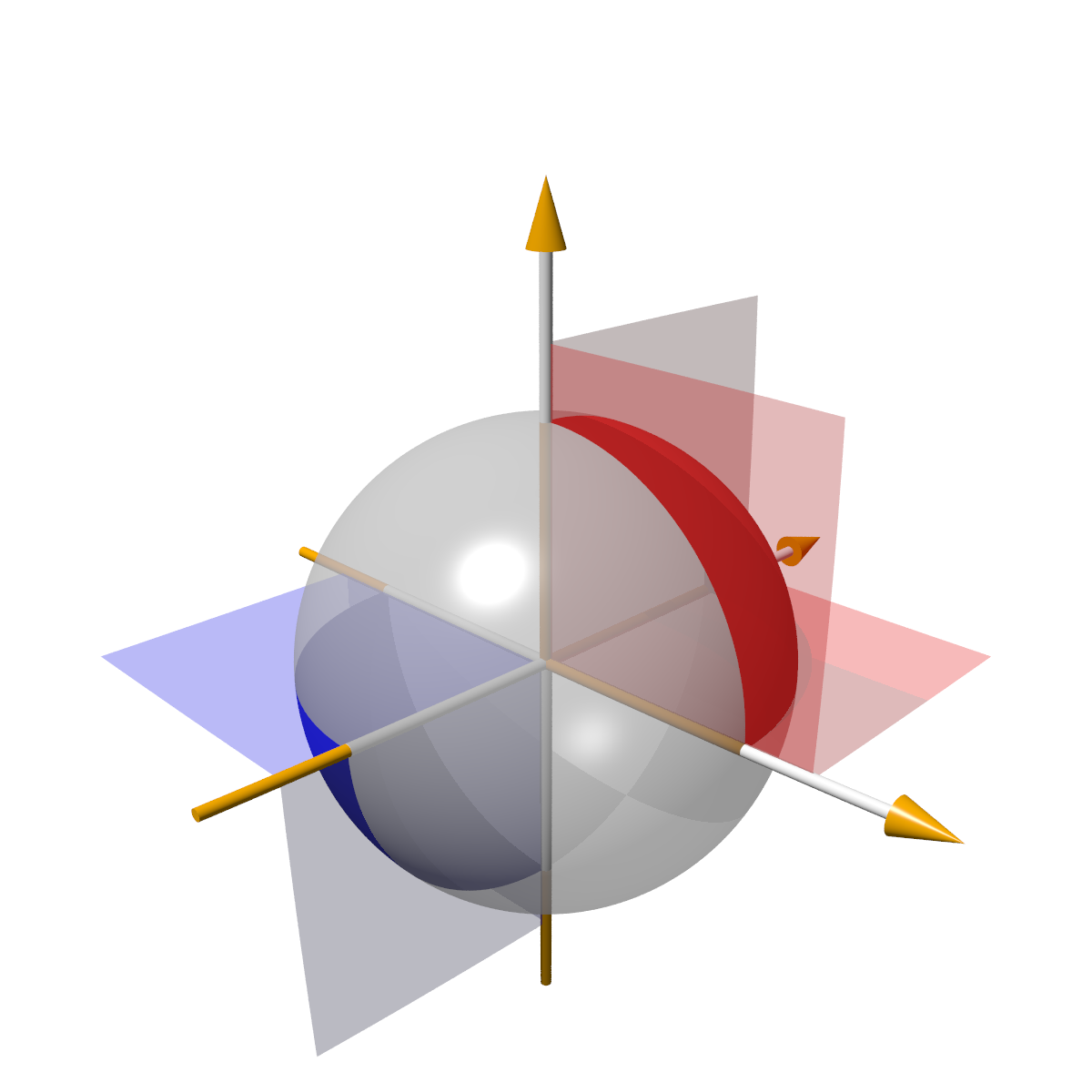}
\begin{picture}(0,0)
\put(-170,152){{\large c)}}
\put(-40,120){${\large \Omega_1}$}
\put(-135,105){${\large \Omega}$}
\put(-150,25){${\large \Omega_2}$}
\put(-63,80){${\large \omega_1}$}
\put(-120,45){${\large \omega_2}$}
\end{picture}
 \includegraphics[width=.4\textwidth]{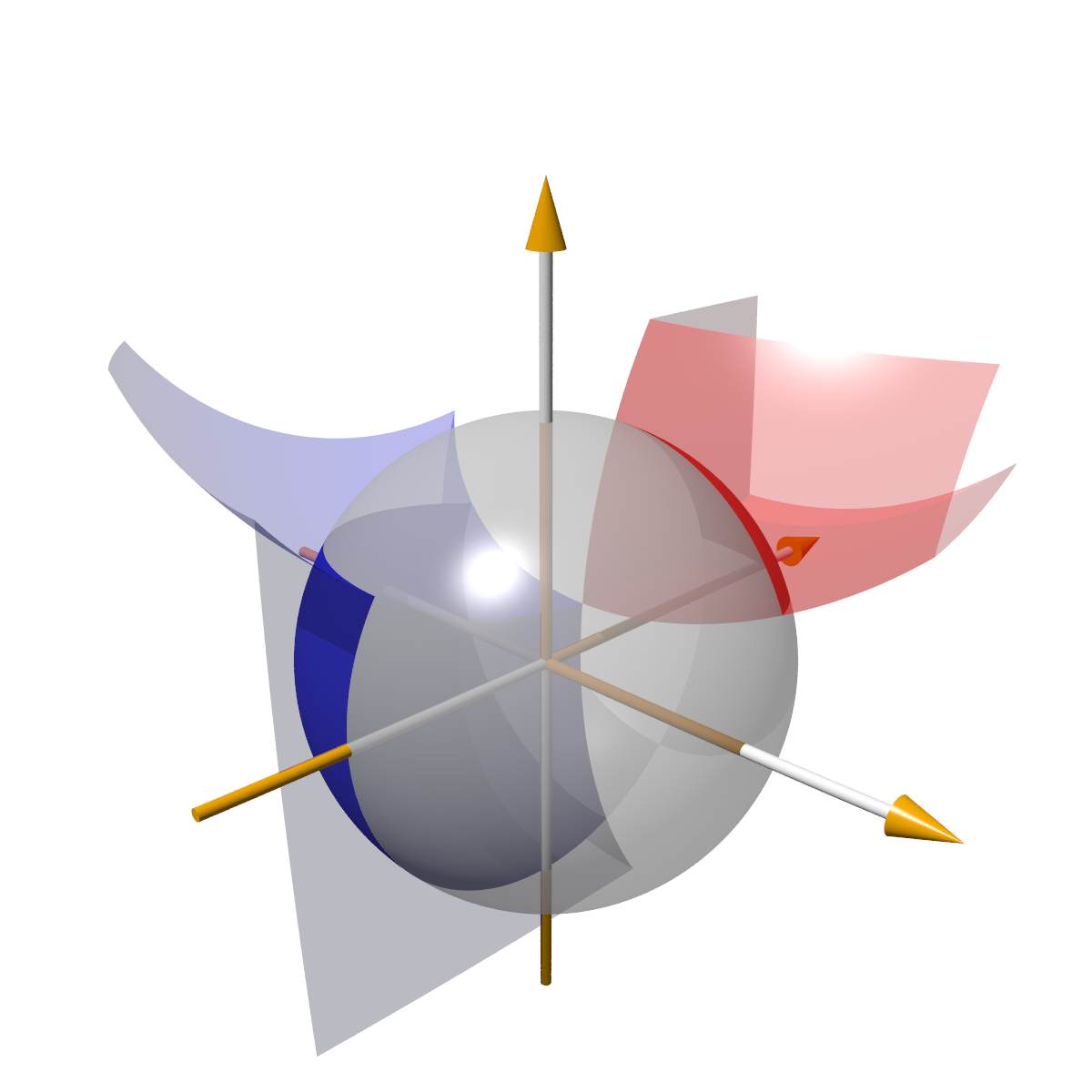}
\begin{picture}(0,0)
\put(-170,152){{\large d)}}
\put(-40,125){${\large \Omega_1}$}
\put(-67,27){${\large \Omega}$}
\put(-150,25){${\large \Omega_2}$}
\put(-67,90){${\large \omega_1}$}
\put(-135,74){${\large \omega_2}$}
\end{picture}
\caption{Geometries related by conformal transformations.
The domain $\Omega$ is the sphere centered
in the origin while the domains $\Omega_i$
are the the ones intersecting $\Omega$ which have, when necessary,
been trimmed to fit the image.
The domains $\omega_i$ are the the darker portions
of $\Omega$.
Panel $a$ refers to the cap geometry ($i$) with
$\theta=0.45\pi/2$ and when it subject to transformation
VIII in table \ref{transformation_table} it is mapped
onto the geometry shown in panel $b$.
Please note that the sphere $\Omega_2$
gets mapped onto the exterior of a sphere
under the action of the conformal transformation.
Panel $c$ is the octant geometry ($ii$)
and when acted on with transformation
V in table \ref{transformation_table}
it is mapped onto the geometry shown in panel $d$.
} \label{fig1}
\end{figure}

We will examine the following percolation geometries:

$i)$ If we take the the domain $\Omega$
to be the unit sphere $x^2 +y^2 + z^2 < 1$ 
and the domains defining
the boundary domains $\Omega_i$, $i=1,2$
to be spheres intersecting with
$\Omega$ at right angles; thus
the problem reduces to the calculation
of crossing among two (disjoint) spherical caps
on a sphere. The starting geometry
(I in our numbering scheme \ref{transformation_table})
is defined by two caps
enclosed by parallels with azimuthal
angles $\theta$ and $\pi-\theta$
as depicted in Figure \ref{fig1}
in panel $a$
together with a geometry obtained
by applying the conformal transformation VIII
in table \ref{transformation_table}.

By use of conformal
transformations it is easy to see that
the crossing probability will depend
only on one anharmonic ratio $\alpha$.
$\alpha$ is easily evaluated
by mapping the surface of the sphere
by a stereographic projection
onto the (extended) plane and having the
caps $\omega_1$ and $\omega_2$ 
mapped onto two circles with center in the origin 
($z_1$) and the point at
infinity ($z_4$). Scale invariance
allows us now to set the radius of $\omega'_1$
to be one ($z_2$). The only free parameter
left is the radius of $\omega'_2$ which we call
$z_3$.
We define the $\alpha$
to be the anharmonic ratio
\begin{equation}
\alpha=\frac{(z_2-z_1)(z_4-z_3)}{(z_3-z_1)(z_4-z_2)} = \tan^2 \left(\frac{\theta}{2}\right).\label{anharm_ratio}
\end{equation}
written with its relation to the angle $\theta$.
In this geometric setting conformal invariance 
implies that $\pi_\times$ depends \emph{only
on $\alpha$}.

$ii)$ The second geometry we will consider
is the crossing probability among two spherical
triangles on the unit sphere. The domains
$\Omega_1$ and $\Omega_2$ will be 
chosen as the octants $x, y, z \geq 0$
and $x, y, z \leq 0$ respectively.
The starting geometry is depicted
in panel $c$ of figure \ref{fig1}.
In panel $d$ of figure \ref{fig1}
we show the octant geometry when
transformed with transformation
V of Table \ref{transformation_table}.


\section{Percolation models}\label{percolation_models}

In order to test the invariance
of crossing probabilities
we will consider two types
of percolation models, a discrete
and a continuum one.

\subsection{Discrete model}
Take a (simple) cubic lattice $\mathbb{L}$ of lattice 
constant $\lambda$ 
i.e. $\mathbb{L} = \lambda (\mathbb{Z}^3+\{1/2,1/2,1/2\})$.
The shift by the vector $\{1/2,1/2,1/2\}$
has been introduced in order
to grant better scaling properties
of our finite lattice realization.
We have specialized to the
site percolation problem.
The sites will be filled 
with a probability $p$.
Such a model has been one of the most widely studied
among the percolation models in three spatial
dimension.
As known the percolation
transition for $p=p_c$
where $p_c=0.31160768(15)$ \cite{xiao_2014}.
As for the realization of the geometries
introduced in Section \ref{percolation}
are concerned in the present model
a point will be considered inside the domain
if it belongs to $\mathbb{L} \cap \Omega$
while the points belonging to $\omega_i$
are the ones in $\mathbb{L} \cap (\Omega_i \setminus \Omega)$
having at least a nearest neighbor in the domain.
In panel $a$ of figure \ref{fig2}
a 2d representation of the above
model.

\subsection{Continuum model}
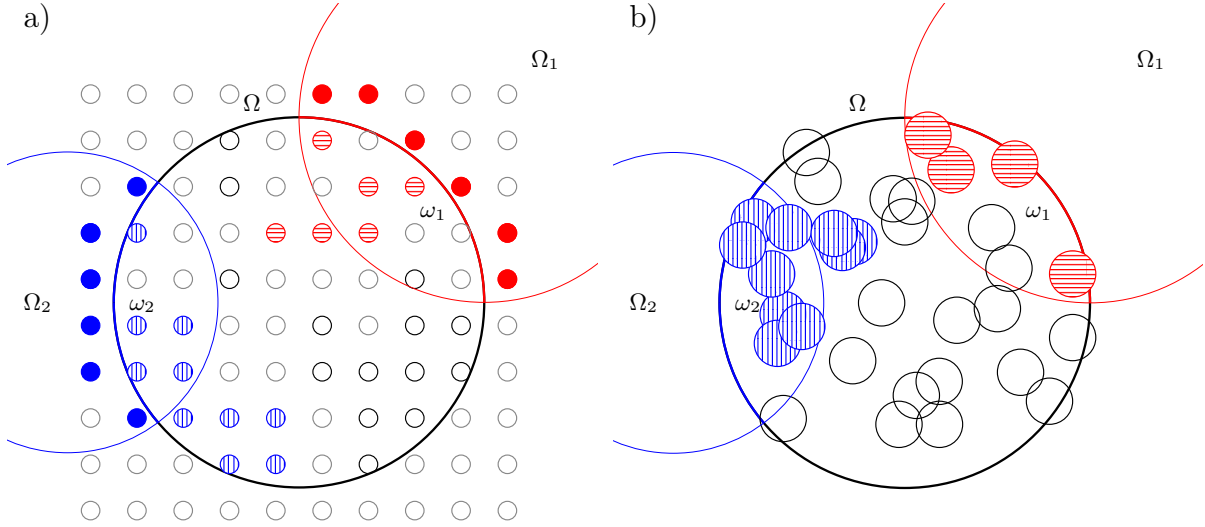
\begin{figure}
\centering
\includegraphics[trim={2.5cm 0 2cm 2cm},clip,width=.49\textwidth]{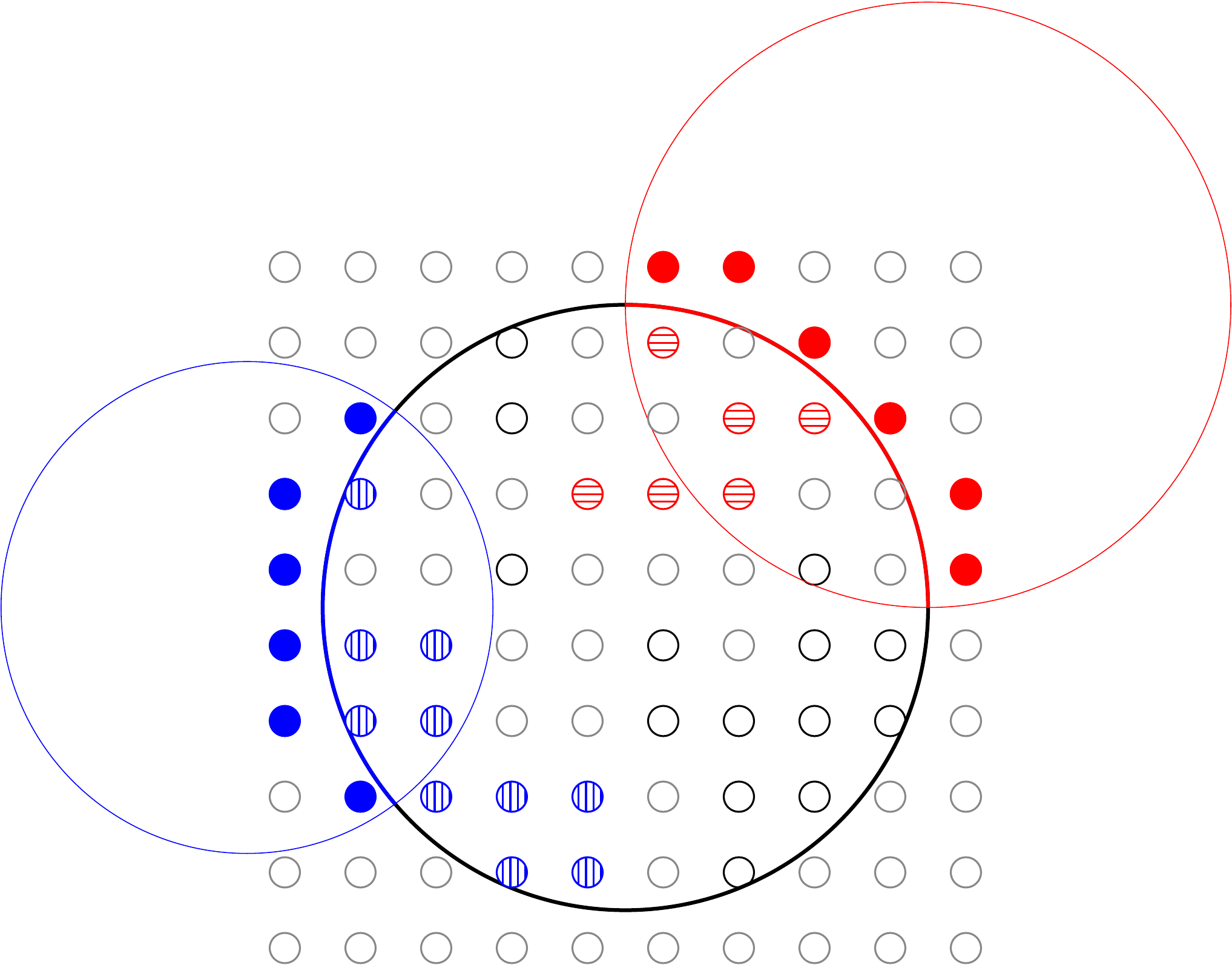}
\begin{picture}(0,0)
\put(-220,187){{\large a)}}
\put(-28,172){${\large \Omega_1}$}
\put(-137,155){${\large \Omega}$}
\put(-220,80){${\large \Omega_2}$}
\put(-70,115){${\large \omega_1}$}
\put(-180,80){${\large \omega_2}$}
\end{picture}
\includegraphics[trim={2.5cm 0 2cm 2cm},clip,width=.49\textwidth]{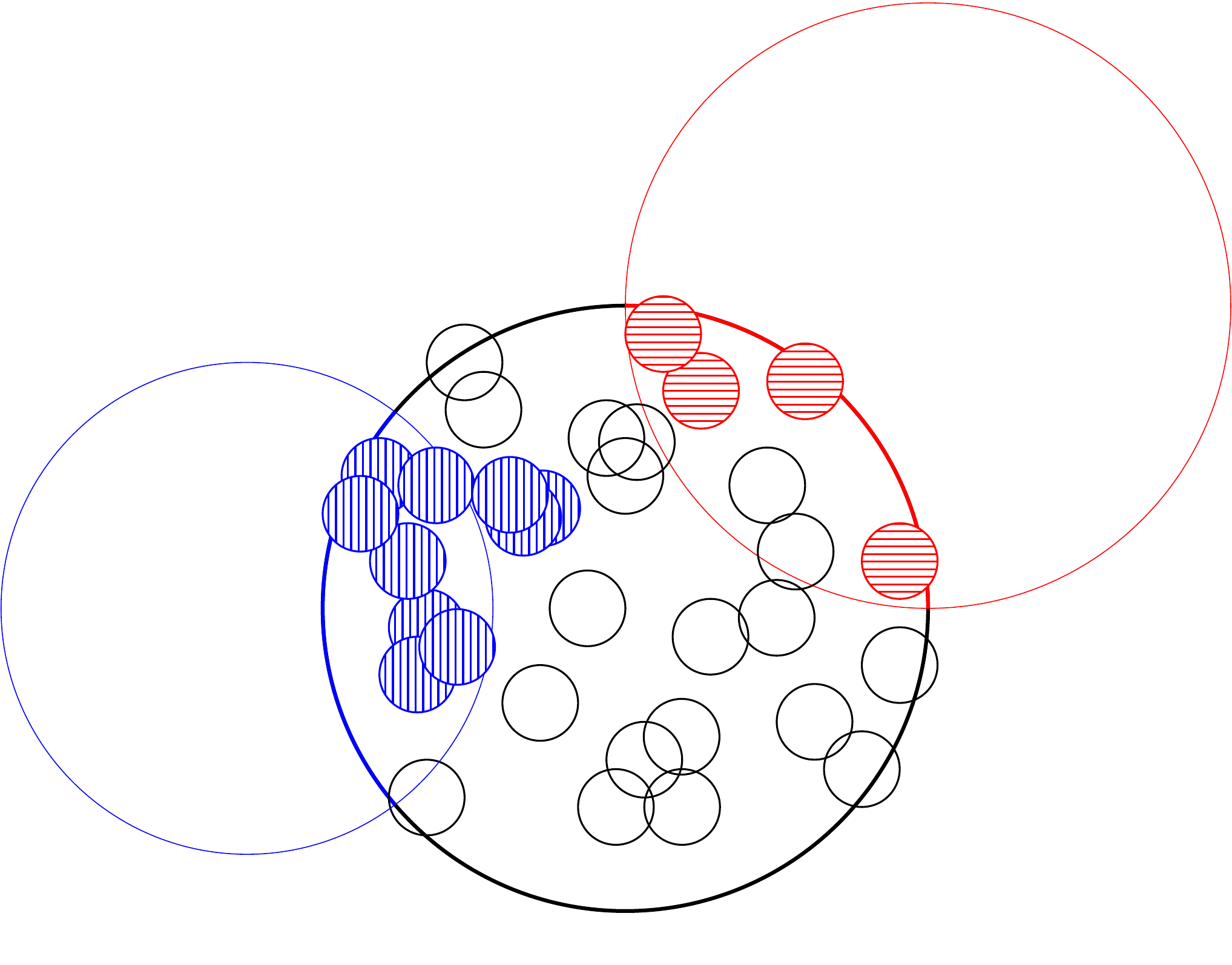}
\begin{picture}(0,0)
\put(-220,187){{\large b)}}
\put(-28,172){${\large \Omega_1}$}
\put(-137,155){${\large \Omega}$}
\put(-220,80){${\large \Omega_2}$}
\put(-70,115){${\large \omega_1}$}
\put(-180,80){${\large \omega_2}$}
\end{picture}
\caption{
2d cartoons of the percolation models we have 
considered in this work. Panel $a$ refers
to the discrete percolation model whilst
panel $b$ to the continuum one.
The shaded circles in $a$ represent occupied
sites, while in $b$ we depict
the randomly occupied circles
(spheres in our 3d model).
The sites (in $a$) and circles (in $b$) with horizontal
and vertical shadings
(red and blue in color) represent
cluster connected to the boundary domains
$\omega_1$ and $\omega_2$ respectively.
In both cases no cluster connecting
$\omega_1$ and $\omega_2$
has been established.
}
\label{fig2}
\end{figure}

The continuum percolation model we have 
considered is the percolation of 
penetrable spheres \cite{mertens_2012}
namely we extract the centers
of the spheres of radius $\rho$ uniformly from the
interior of $\Omega$ (Poisson 
point process). A sphere will belong to the $\omega_i$
if the distance of its center $c$ to $\Omega$
is less or equal than $\rho$ and $c$ is
an element of $\Omega_i$. Please see
panel $b$ of figure \ref{fig2}.

The existence of a percolating cluster
depends on the so called filling factor
$\eta$ defined as the mean number
of objects $n$ times the ratio
of the volume of the filling 
spheres to the total volume
which in our case reads $\eta=n \rho^3$.
The value $\eta_c$ for which the transition
occurs is known from numerical
experiments to be $\eta_c=0.34189(2)$
\cite{torquato_2012} and $\eta_c=0.341888(3)$ \cite{lorenz_2001}
which is to date the most precise estimate
of the critical filling.

\section{Numerical Simulations}
The discrete percolation experiments have been carried out
on a simple cubic lattice $\lambda$ of decreasing lattice
spacing, namely $1/\lambda=8,
16, 32, 64, 128$ in order to perform 
a scaling analysis.
We have employed the Newman-Ziff \cite{newman_2000, newman_2001} algorithm
to calculate the crossing probability.
We have thus obtained the crossing
probability $\pi_\times (n)$ as the lattice
is filled (incrementally) 
with $n$ sites.
From the above probabilities
at fixed filling we can reconstruct
the crossing probability $\pi_\times (p)$ in terms
of \emph{any} occupation probability $p$
via the following binomial convolution:
\begin{equation}
 \pi_\times (p) = \sum_{n=0}^{N} \pi_\times (n) p^n (1-p)^{N-n} 
\binom{N}{n}
\end{equation}

For the continuum percolation experiments
we used spheres of radius $\rho$.
We have considered the sizes of $\rho$
to be$1/\rho = 2^{3}, 2^{3.5}, 2^4, \ldots, 2^{6.5}, 2^{7}$
with centers extracted uniformly from the 
volume under consideration. 

In this case we resorted
to the continuum version of the Newman-Ziff
algorithm detailed in \cite{mertens_2012}.
In this case the measurement 
of crossing probabilities 
with a given number $n$ of spheres
$\pi_\times(n)$ allows to
obtain the crossing probability
for any filling factor $\eta$
by a poissonian convolution:

\begin{equation}
 \pi_\times (\eta) = e^{n/\rho^3}\sum_{n=0}^{\infty} \pi_\times(n) \frac{(n/\rho^3)^n}{n!}
\end{equation}

%
The number of realizations used for
our simulation consists
of up to $10^7$ samples allowing us to
obtain the desired statistical
uncertainties.

For the random number generation we resorted
to the, well tested for percolation
simulations \cite{lee_2007, lee_2008}, Mersenne
twister MT19937 by Matsumoto
and Nishimura \cite{matsumoto_1998}.

\section{Results}

\subsection{Site percolation}

\begin{figure}
\centering
\includegraphics[width=.49\textwidth]{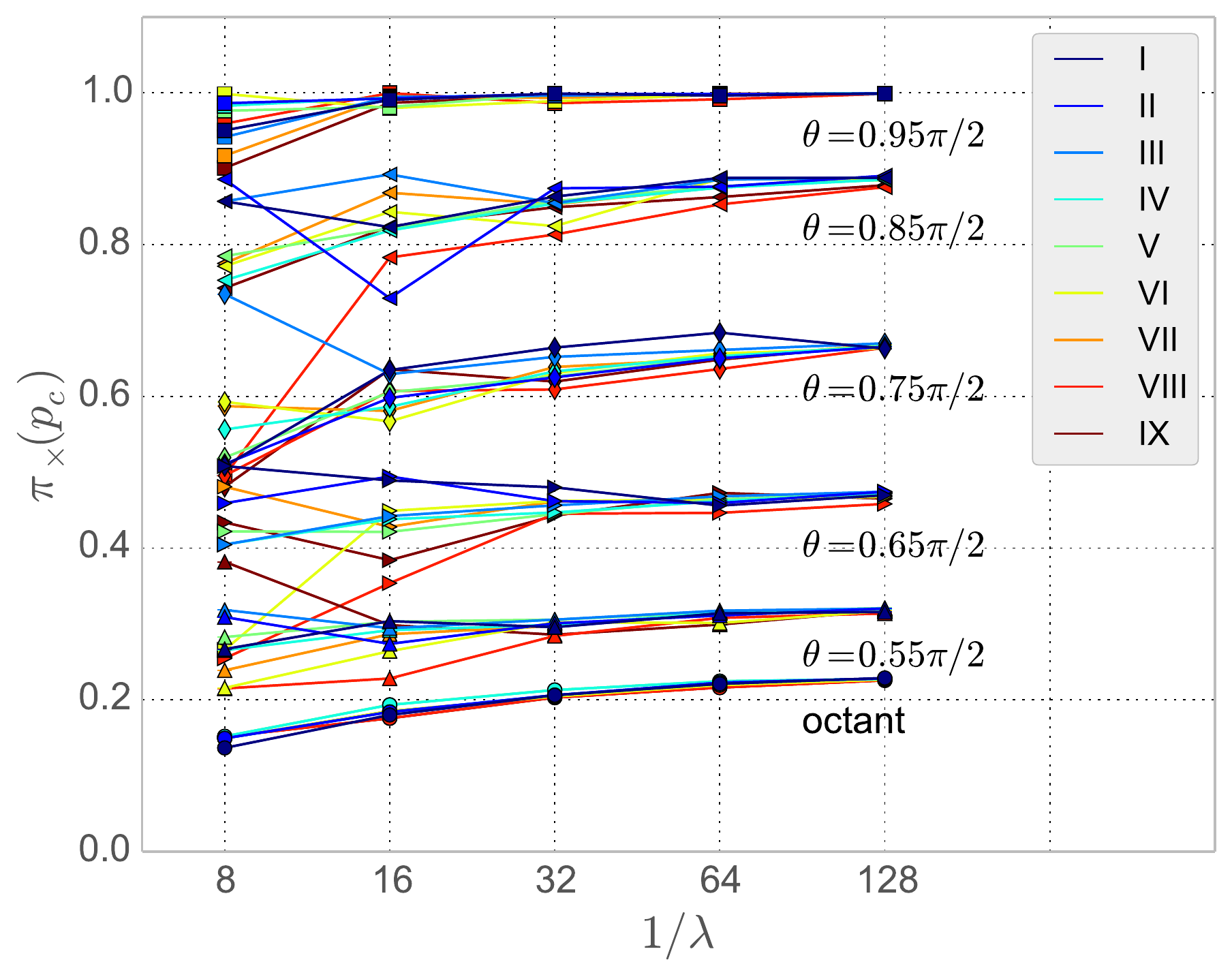}
\includegraphics[width=.49\textwidth]{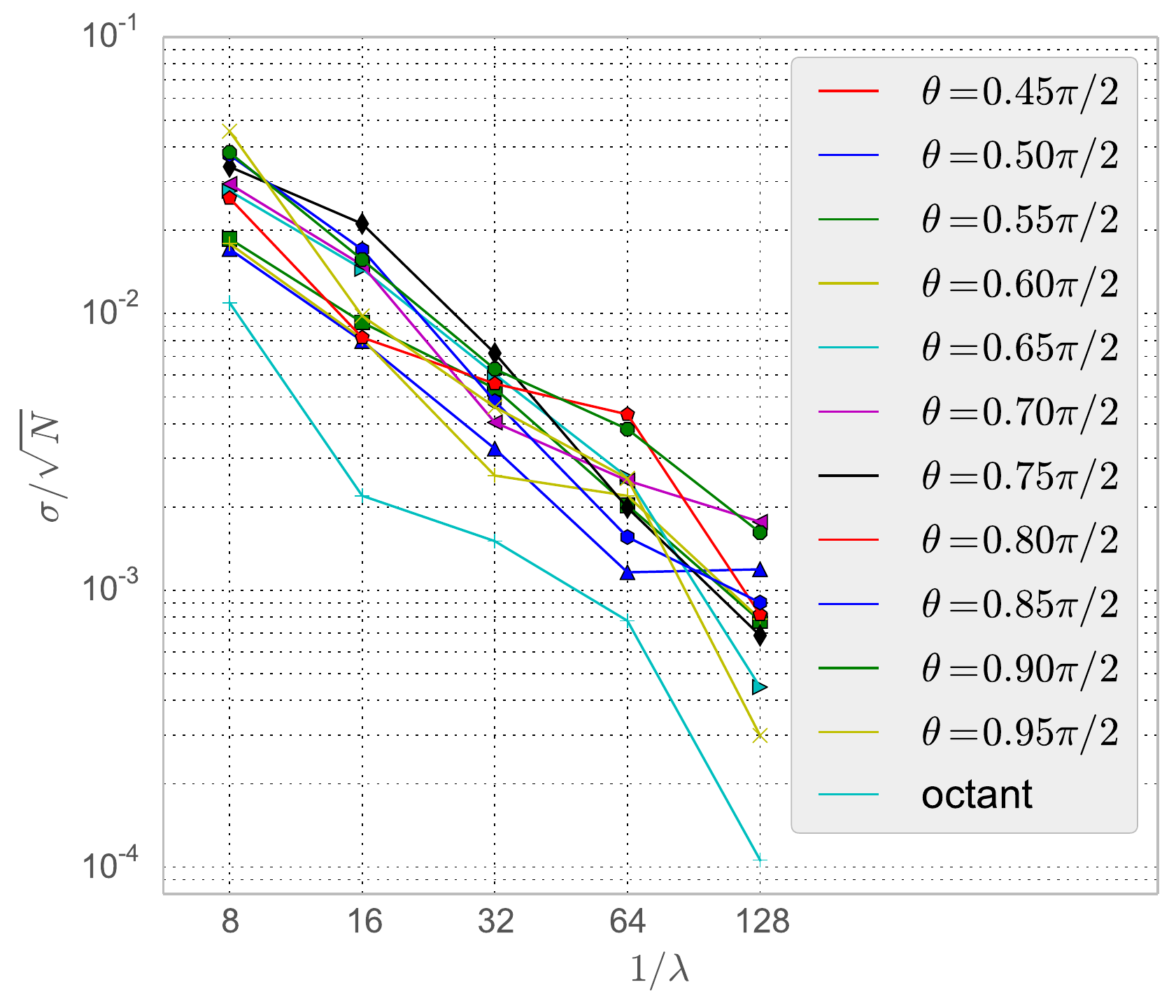}
\caption{(Left panel) Crossing probabilities
at $p_c$ as a function of the 
inverse lattice spacing $1/\lambda$.
The different colors refer to
geometries obtained by applying
the transformations in table
\ref{transformation_table}.
The distinct bunches of curves
refer to different values of
$\theta$ and for the octant
geometry.
Statistical errors are smaller than
the symbols. (Right panel) Variance
of the crossing probabilities
as a function of the system size
for different values of $\theta$.
The lines are just guide to the eyes.
}\label{fig3}
\end{figure}

In figure \ref{fig3} (left panel) we depict the crossing
probability as a function of the inverse
lattice spacing for the cap geometry
for some of the values of $\theta$ considered
and for the octant geometry.
These have been obtained by setting
the occupation probability $p$
to the best known value $p_c=0.31160768(15)$ \cite{xiao_2014}.

As we can see groups of curves corresponding
to the geometries related by a conformal
transformation converge to the same value
as the lattice realization becomes finer. 
Among the various mapped geometries
we observe that VIII and IX are the ones
differing the most from the others.
This is to be expected since they are
subject to more severe deformation
such that one of the caps becomes
very small (cfr. panel $b$ of Figure \ref{fig1}
referring to a geometry obtained
by applying transformation VIII). 
When this happens
even the finest simulated finite lattice realizations
have stronger finite size effects.

In order to quantitatively assess the
convergence to the same value 
in \ref{fig3} (right panel) we plot the variance of
$\pi_\times$ for the nine geometries
considered for different values of 
$\theta$ which decay to zero, signalling
the presence of the conformal symmetry
in the thermodynamic limit.

\begin{figure}
\centering
\includegraphics[width=.49\textwidth]{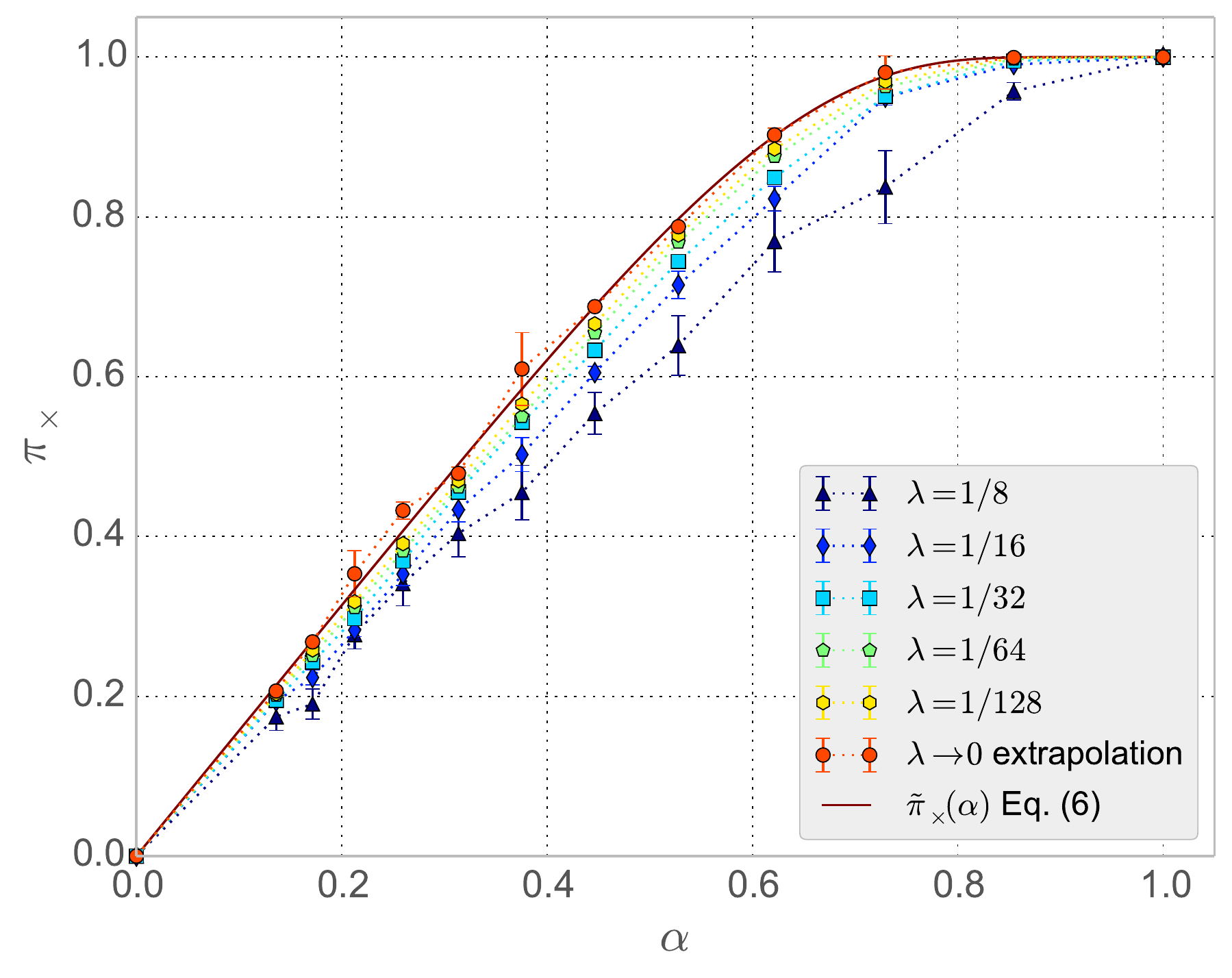}
\caption{Crossing function
for two spherical caps
on a sphere in term of the
anharmonic ratio $\alpha$ (defined in
\ref{anharm_ratio}).
The different curves refer
to different system sizes, and to
the extrapolation to the infinite
size system. The dotted lines are just guide to the eyes.
The continuous line refers to
the function $\tilde{\pi}(\alpha)$
defined in Equation \eqref{guessed_function}.}\label{fig4}
\end{figure}

By extrapolating these finite size results
to the thermodynamic limit allows us
to obtain a numerical estimate of the
crossing probabilities for any two
spherical cap on the sphere.
The crossing probability in terms 
of the anharmonic ratio $\alpha$
is depicted in \ref{fig4}.
In this figure we also plot
the function 

\begin{equation}
\tilde{\pi}_\times(\alpha)=\tanh \left(\tan \frac{\alpha \pi}{2} \right)\label{guessed_function}
\end{equation}

which appears to describe 
very well the values extrapolated
in the thermodynamic limit.
Indeed fitting the extrapolated values
the function $\tanh \left( a \tan  \frac{\alpha \pi}{2} \right)$
with $a$ free parameter gives
a value of $a=0.989(6)$ very close to 1.

The function $\pi_\times$ 
we have numerically estimated
and for which we have given 
an analytic approximation
$\tilde{\pi}_\times(\alpha)$
is somewhat analogous to the 
Cardy's formula \cite{cardy_1994} 
although in the two dimensional 
case, due to Riemann
mapping theorem, all of the crossing
probabilities among two distinct segment
on the boundary of a connected
domain can be obtained by this formula.
In the three dimensional case
instead the knowledge of the 
function plotted in \ref{fig4} 
allows us to calculate
the crossing probability
among two arbitrary spherical 
caps on the surface of a sphere.

\subsection{Continuum percolation}

\begin{figure}
\centering
\includegraphics[width=.49\textwidth]{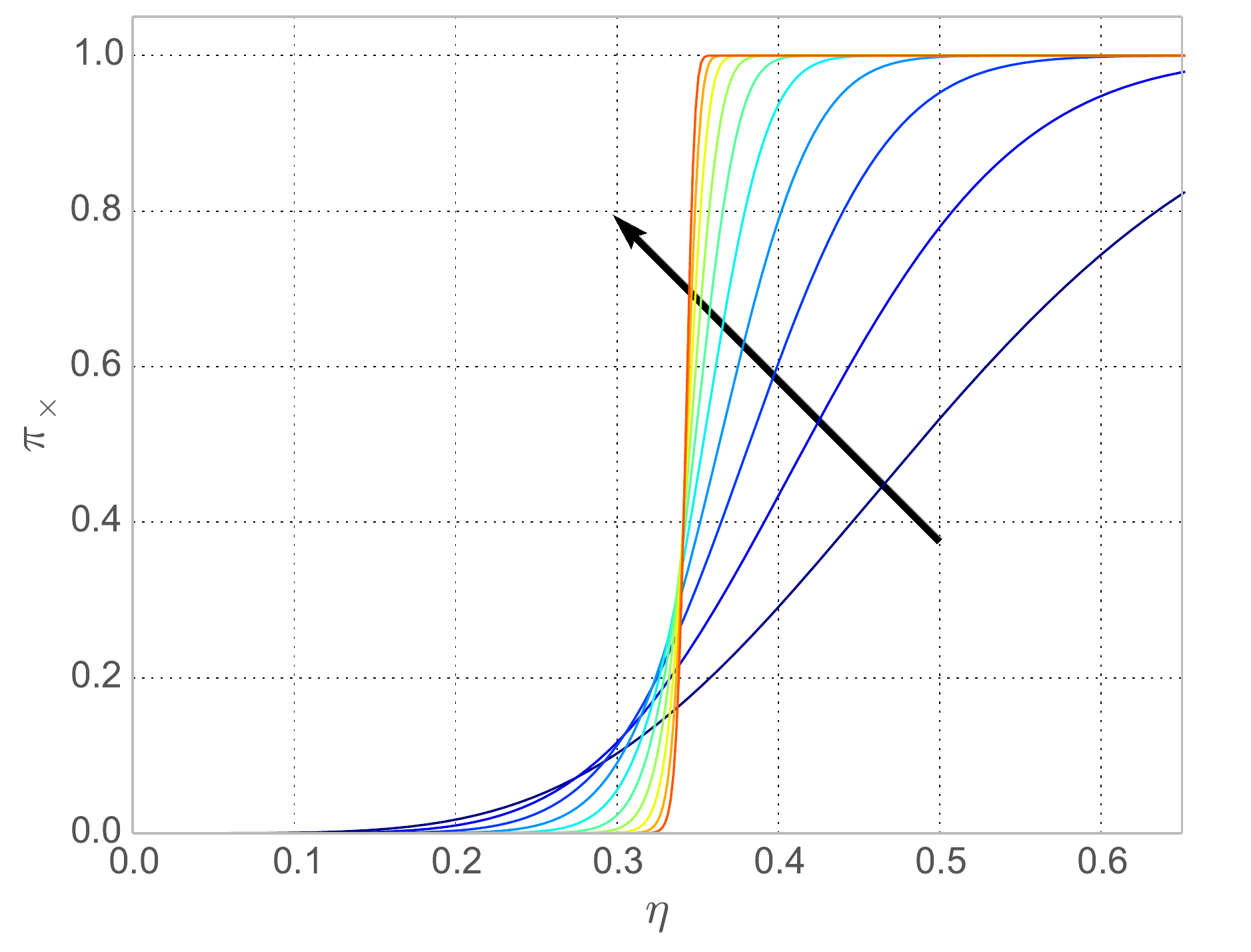}
\includegraphics[width=.49\textwidth]{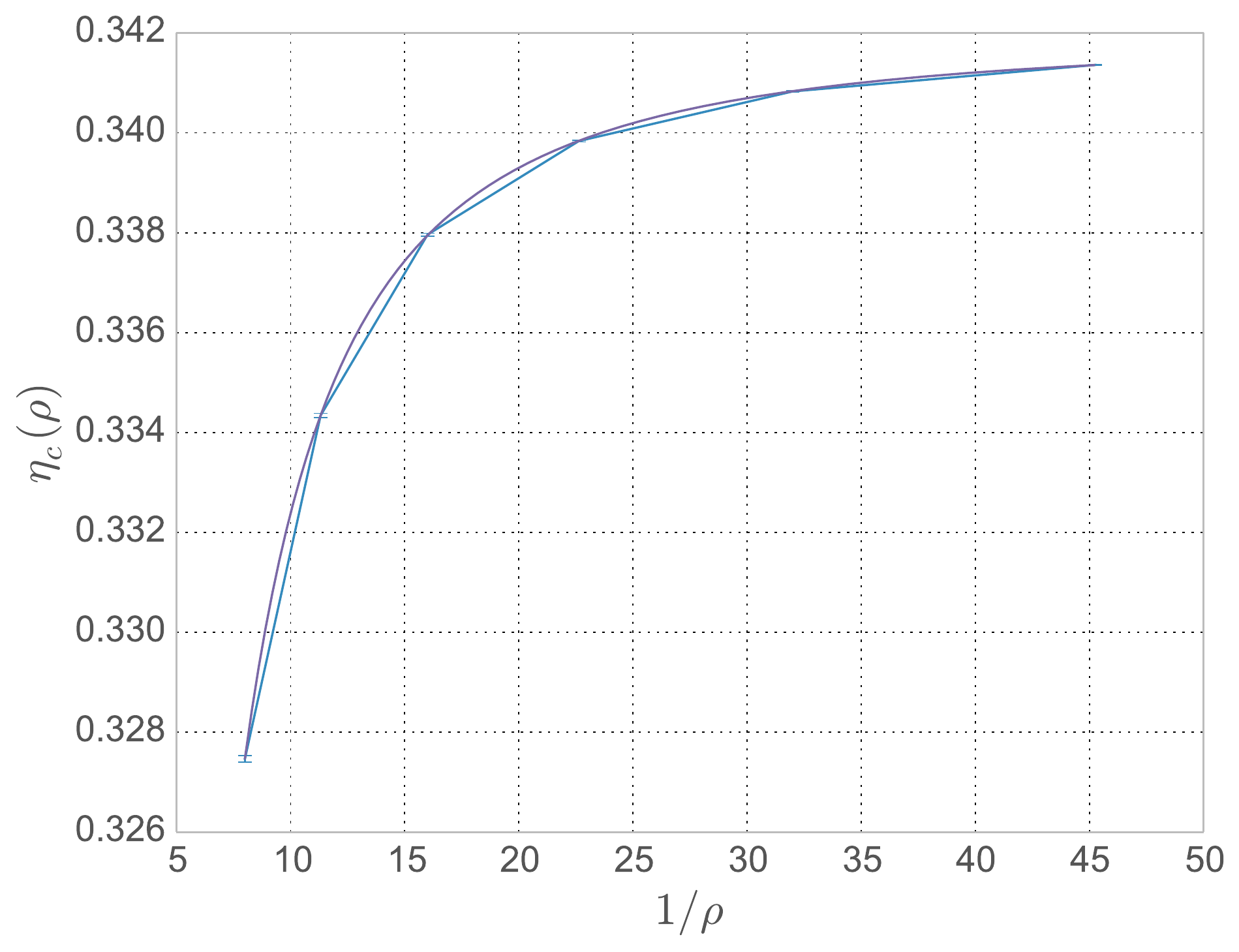}
\caption{(Left panel)Crossing probability as a function 
of the filling factor for the cap geometry 
with $\theta=0.65 \pi/2$. The different lines 
refer to the filling spheres of increasing radius 
$1/\rho=2^{2},2^{2.5},\ldots, 2^{6.5}$
in the direction of the arrow.
The errors are not showed being smaller than $2.5\cdot 10^{-4}$
for all the points i.e. smaller than the lines size.
(Right panel) Finite size estimations of $\eta_c(\rho)$
as a function of $1/\rho$. The continuous line
is the best fitting function $0.341935 - 0.687988 \rho^{1.85758}$.
The two lowest values of $1/\rho=2^{2},2^{2.5}$
have been excluded from the fit.
}
\label{fig5}
\end{figure}

In contrast to the discrete model the continuum
one has the added benefit of having a continuously tunable 
parameter $\rho$.
Moreover as we lower $\rho$ 
the various crossing probabilities
approach the thermodynamic
limit in a much smoother
way.
In \ref{fig5} we plot the crossing probability
as function of the filling fraction
for decreasing filling sphere radii
for $\theta=0.75 \pi/2$ for the
configuration I. These curves clearly develop 
a steep profile as we approach the thermodynamic limit
as shown in panel $a$ of Figure \ref{fig5}. 
We can use them to obtain reliable
estimates of the critical filling
fraction. In fact, by defining
a size dependent $\eta_c(\rho)$
(the so called cell-to-cell
estimator \cite{reynolds_1980}) 
as the value for which the curves
meet:
\begin{equation}
 \pi_\times^\rho (\eta_c(\rho)) =  \pi_\times^{\rho/2} (\eta_c(\rho))
\end{equation}
we can easily obtain an estimate for
$\eta_c$. This
analysis, shown in panel $b$ of
Figure \ref{fig5}, with
the $\theta=0.65\pi/2$ cap geometry
I leads us to estimate the critical $\eta$ as 
$\eta_c=0.341935(8)$.
We have chosen the value
$\theta=0.65\pi/2$
and the configuration
I since, among the many geometries
we have simulated, it
represents ``best'' the thermodynamic
limit for it has the
domains $\omega_1$ and $\omega_2$ 
of the same size and
the ratio of sum the total area of the
patches $\omega_i$ and the remaining
surface is closest to 1.
This estimate has to be compare with the ones available in
literature \cite{torquato_2012} and \cite{lorenz_2001}
which provide values of
$\eta_c=0.34189(2)$ and $\eta_c=0.341888(3)$
respectively. Those results were obtained with growth 
algorithms on large systems. In our simulation the lowest value 
of $\theta$ ($\theta=0.45\pi/2$) simulated
cap geometry resembles somehow the study
of percolating clusters
connecting two sides of a cube
since the ratio of patches area
to open surface area is closest
to $1/2$.
Performing the analysis on this
configuration we obtain
a value of $\eta_c=0.34190(1)$
which is consistent with the above cited values,
showing how the determination of $\eta_c$
sensibly depends on the geometry 
for the sizes we have examined.
We will rely on our ``best'' determined 
estimate $\eta_c=0.341935(8)$ for the
subsequent analysis but we anticipate that
the final results will not be crucially
affected by the choice of $\eta_c$.

\begin{figure}
\centering
\includegraphics[width=.49\textwidth]{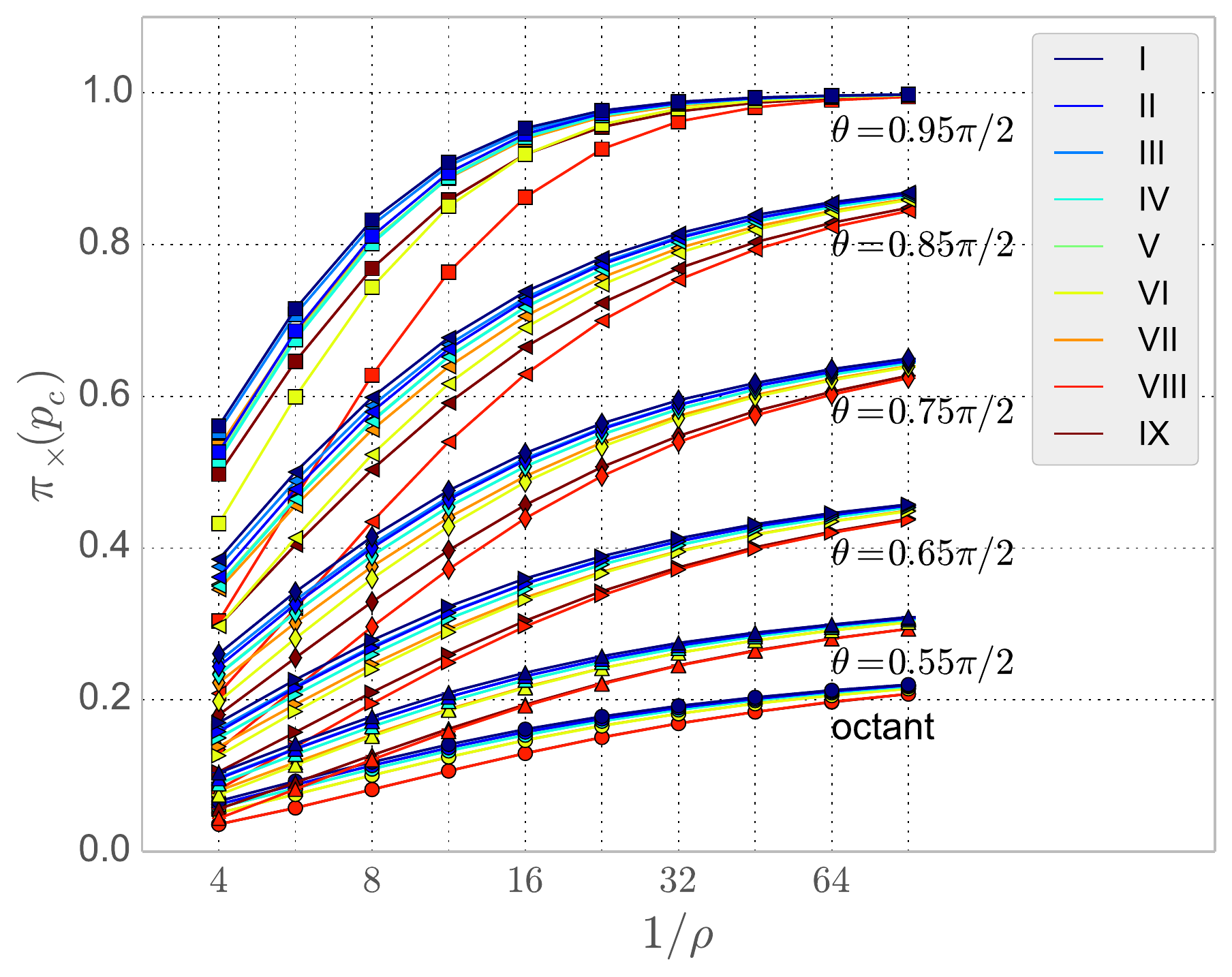}
\includegraphics[width=.49\textwidth]{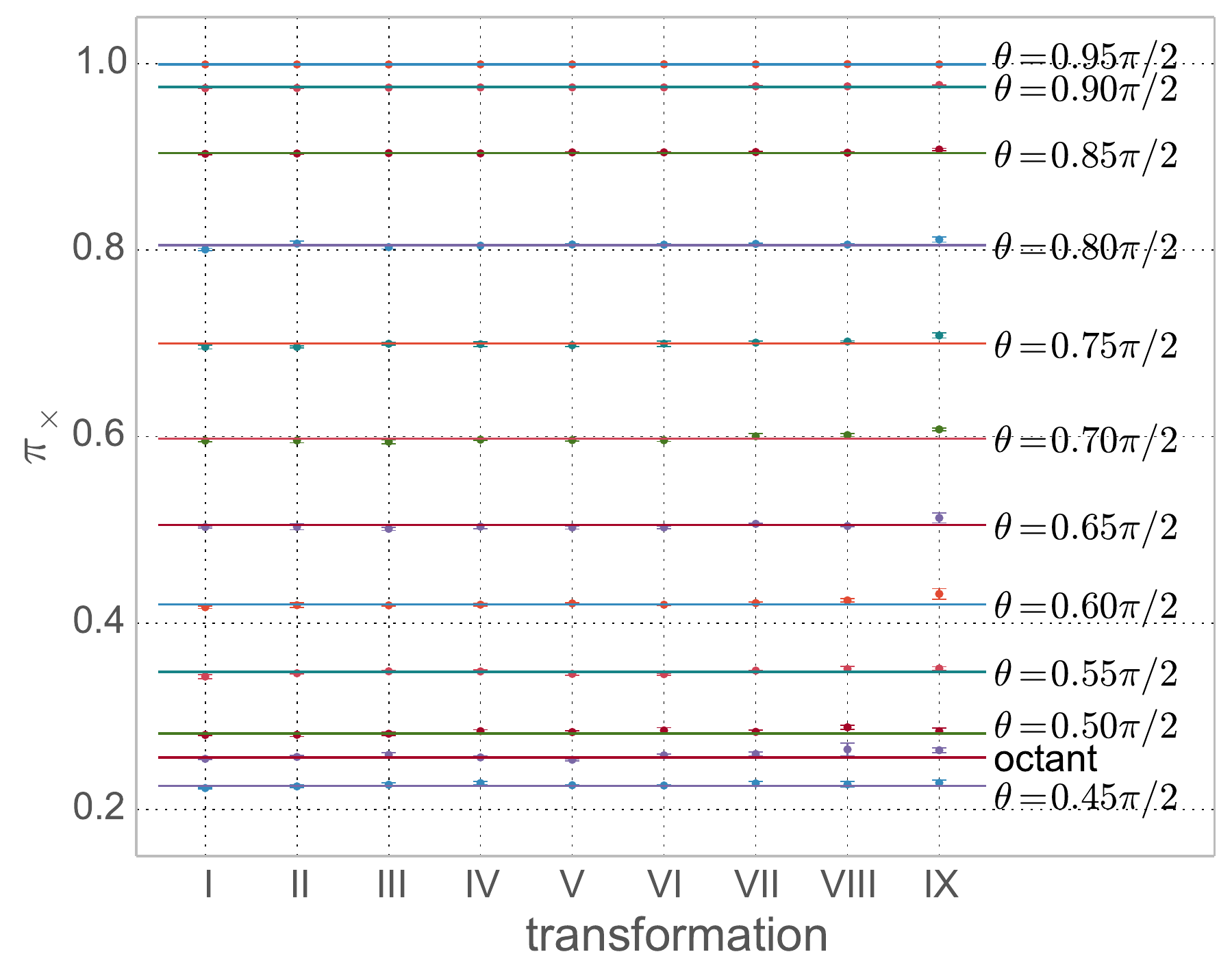}
\caption{(Left panel) Crossing probabilities
$\pi_\times(L)$ of conformally equivalent cap geometries 
and octant geometry as the radius $\rho$ of the
filling spheres is reduced, cfr.
left panel of Figure \ref{fig3}. The crossing
probabilies are calculated at $\eta=\eta_c$.
(Right panel) Extrapolated values of 
$\pi_\times$ in the thermodynamic
limit for the caps and octant geometry with
for the different
conformal transformations considered
\ref{transformation_table}.
}
\label{fig6}
\end{figure}

The convergence to the same value
for $\pi_\times$ for geometries
related by a conformal transformation 
can seen in the left panel of Figure
\ref{fig6} which is the continuum analogue of 
the left panel in Figure \ref{fig3}.
The curves for $\pi_\times$
nicely converge to the thermodynamic
limit, since the effects from discretization
are not present; on the other hand finite
size effects are stronger, for the range
of $\rho$ simulated, since their distance
is in general bigger than the
curves for the discrete systems.
Once again we observe that configurations
generated with VIII and IX transformations
are more distant from the other
curves confirming that, for the same
value of $\rho$, they are more far
from the thermodynamic limit.

The good behavior of the finite
realizations allows to have 
an independent extrapolation for each configuration.
This is shown in the right panel of Figure
\ref{fig6} where the extrapolated
values of the crossing probabilities
are shown for the set of conformal
transformations examined together
with the mean over the various
configurations.
This confirms the onset of conformal
invariance even for a continuum percolation model

\begin{figure}
\centering
\includegraphics[width=.49\textwidth]{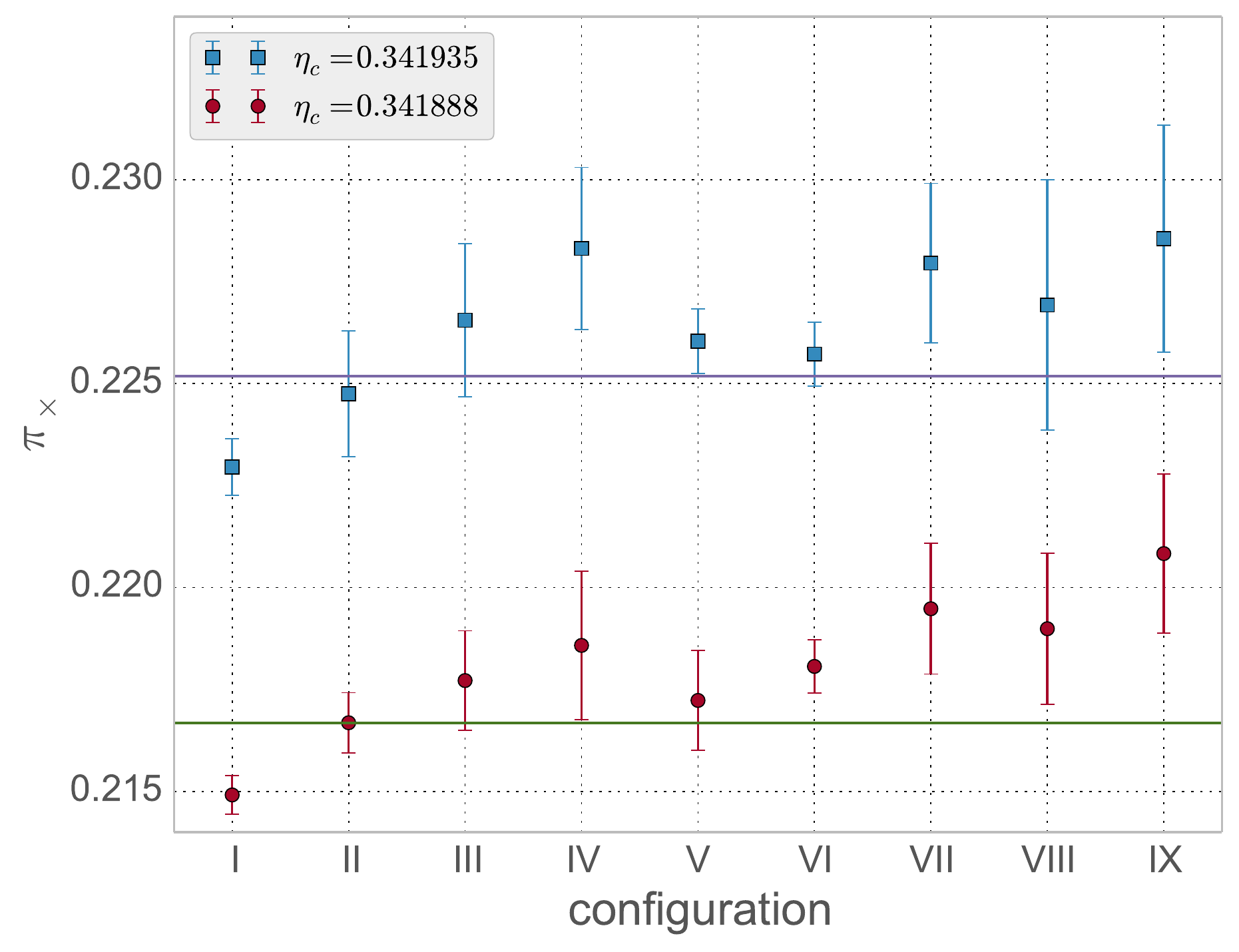}
\includegraphics[width=.49\textwidth]{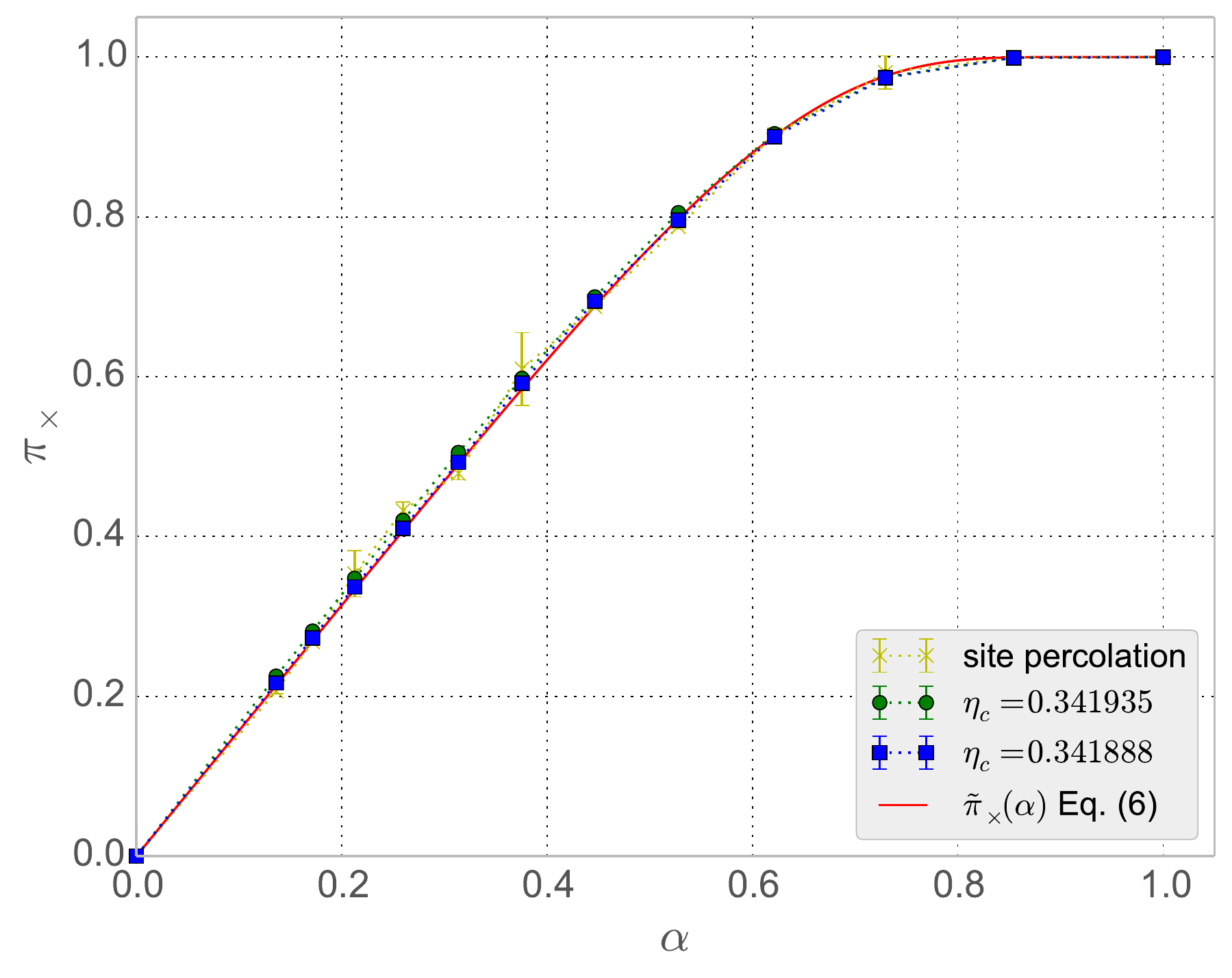}
\caption{(Left panel) Extrapolated values of 
$\pi_\times$ in the thermodynamic
limit for the cap geometry with
$\theta=0.45\pi/2$ for the different
conformal transformations considered
\ref{transformation_table}.
The two curves refer to the values
obtained setting $\eta$ to our best
estimated value (upper curve) and the value.
given in \cite{lorenz_2001}.
(Right panel) Comparison between 
the values extrapolated at infinity
for the site percolation model and the 
continuum percolation model with the two estimates
for $\eta_c$ as a function of the anharmonic
ratio $\alpha$ \eqref{anharm_ratio}. 
We also report the function $\tilde\pi(\alpha)$
defined in \eqref{guessed_function}.
}\label{fig7}
\end{figure}

For comparison we also report, in the
left panel of figure \ref{fig7}, the same analysis
performed with the critical filling
reported in \cite{lorenz_2001}
for the cap geometry with $\theta=0.45\pi/2$.
As we can see in both cases 
the different transformed geometries
give the same value within
$1.5 \sigma$ so they are consistent
with each other. 
This proves that within
our error, both values of $\eta_c$
give the same value for the crossing
probability in domains related
by a conformal mapping
although the value of $\pi_\times$
depends on the chosen $\eta_c$.
Similar findings are obtained for
the other values of $\theta$ examined
and for the octant geometry.

The function $\pi(\alpha)$ \eqref{guessed_function}, introduced
in the previous section, ruling
the crossing probability among two caps
in on the sphere, is depicted in
the left panel of \ref{fig7}. As we can see the values
are consistent, within errors, with the
function obtained from the site percolation
model. This provides an evidence
of universality of $\pi_\times$
in percolation. Moreover if 
we compare it with the analytic
function $\tilde{\pi}_\times$ \ref{guessed_function}
proposed in the previous section
we find an excellent agreement
with the data obtained setting
$\eta_c$ to the value provided in
\cite{lorenz_2001} while 
if we choose our best estimate
of $\eta_c$ we have a less faithful
representation of the numerical data 
by \ref{guessed_function}.
If we try to fit the above data
with the more general function 
$\tanh \left( a \tan  \frac{\alpha \pi}{2} \right)$
indeed we obtain the estimates of 
$a=1.022(6)$ for $\eta_c=0.341935$ (our best value)
and $a=1.005(5)$ for $\eta_c=0.341888$ (given in \cite{lorenz_2001}).

\section{Conclusions}
By means of numerical experiments
we have assessed the invariance
under conformal transformation
of selected both discrete 
and continuous percolation problems 
in three dimensions at criticality in bounded domains. 
We also proposed 
an analytical function approximating
with very good accuracy
the crossing probability
among two arbitrary
spherical caps on a sphere.

We hope that our work 
can stimulate from one side a study of percolation by conformal bootstrap techniques  
and from the other further investigation of symmetries of general percolation models. 
Future work in our opinion interesting will entail the 
analysis of bulk observables and the study of different
statistical mechanics models such as $O(N)$ and Potts models.\\
 
{\em Note added:} During final stage of this work the
authors noticed on \verb|arXiv| a very recent and very interesting 
paper on the numerical investigation
of Ising model on spherical domains
with the help of the insight gained 
from the conformal bootstrap
analysis of the model \cite{penedones_2015}
for the finite size scaling analysis
of correlators.
The work \cite{penedones_2015} concentrates on 
observables lying in the bulk
of a three dimensional Ising model.
It would be interesting to
extend the analysis of \cite{penedones_2015}
to observables living on the
boundary.
For the percolation the corresponding
analysis of operators living
either in the bulk or in the boundary 
of the pertinent three dimensional boundary
CFT has, to the best of our
knowledge, not been derived.

{\em Acknowledgements:} 
We acknowledge useful discussions
with Defenu~N and Delfino~G.
The authors 
benefitted from computational resources
from the Iscra C project COSY3D
at CINECA Bologna Italy.

\end{document}